\documentclass[aps,twocolumn,prd,superscriptaddress,preprintnumbers,showpacs]{revtex4}
\usepackage{graphicx}

\newcommand{\be}{\begin{equation}}
\newcommand{\ee}{\end{equation}}
\newcommand{\bea}{\begin{eqnarray}}
\newcommand{\eea}{\end{eqnarray}}

\begin{document}

\preprint{}

\title{The gravitational wave background from super-inflation in Loop
  Quantum Cosmology}
\author{E. J. Copeland}
\email[]{Ed.Copeland@nottingham.ac.uk}
\affiliation{School of Physics and Astronomy, University of Nottingham, University Park, Nottingham NG7 2RD, UK}
\author{D. J. Mulryne}
\email[]{D.Mulryne@damtp.cam.ac.uk}
\affiliation{Department of Applied Mathematics and Theoretical Physics, Wilberforce Road, Cambridge, CB3 0WA, UK}
\author{N. J. Nunes}
\email[]{nunes@damtp.cam.ac.uk}
\affiliation{Department of Applied Mathematics and Theoretical Physics, Wilberforce Road, Cambridge, CB3 0WA, UK}
\author{M. Shaeri}
\email[]{ppxms1@nottingham.ac.uk }
\affiliation{School of Physics and Astronomy, University of Nottingham, University Park, Nottingham NG7 2RD, UK}

\date{\today}

\begin{abstract}
 We investigate the behaviour of tensor fluctuations in Loop Quantum Cosmology, 
focusing on a class of scaling solutions which admit a near scale-invariant scalar field power spectrum.
We obtain the spectral index of the gravitational field perturbations, and find 
a strong blue tilt in the power spectrum with $n_t \approx 2$. The amplitude of tensor 
modes are, therefore, suppressed by many orders of magnitude on large scales compared to those 
predicted by the standard inflationary scenario where $n_t \approx 0$.

\end{abstract}

\maketitle

\section{Introduction }

A key result of the inflationary scenario of the very 
early universe \cite{inflation}  is that it gives rise to a stochastic background of gravitational wave 
radiation (tensor perturbations) \cite{stochastic-gw}. Such a background is in principle observable and could be 
used to distinguish between different models of inflation \cite{distinguish}. Alternative proposes 
such as the ekpyrotic/cyclic scenario \cite{ekpyrotic, Boyle:2003km} 
or phantom super-inflation \cite{phantomInflation,Piao:2006jz} 
lead to very different predictions for the spectral tilt 
of tensor perturbations when compared with the simplest inflationary 
models \cite{Boyle:2003km}, implying the gravitational wave background  could also 
be a powerful discriminator between competing theories. In this work we will be concerned with the 
gravitational wave background produced by super-inflationary scenarios within Loop Quantum Cosmology (LQC).

Loop Quantum Gravity (LQG) \cite{LQG} is  a background independent and 
non-perturbative canonical 
quantisation of general relativity based on Ashtekar variables:
su(2) valued connections
and conjugate triads.  The variables used in
the quantisation scheme are holonomies of the connection and
fluxes of the triad. Restriction to symmetric states followed by quantisation using the 
techniques of LQG gives rise to LQC \cite{LQC}. 
Although much regarding LQC is not fully understood, in particular its relation to full LQG, 
it has produced a number of intriguing results  and resolved the 
problem of singular evolution present in the earlier Wheeler de Witt approach to quantum cosmology \cite{NS}. 

One particularly useful approach 
to studying the early universe in the context of LQC is that of deriving and studying effective 
equations of motion \cite{rho2, inverse,Bojowald:2002ny,Bojowald:2002nz,Vandersloot:2005kh}. 
In the context of inflation in LQC many intriguing effects have been investigated in this setting 
\cite{Bojowald:2002nz,LQCinflation,Mulryne:2006cz, Copeland:2007qt}.
While such equations cannot probe fully quantum regimes, they nevertheless 
incorporate quantum modifications into classical evolution equations whilst avoiding the interpretational 
difficulties inherent in fully quantum equations. In isotropic settings 
two sets of modifications have been predominately considered (see 
Ref. \cite{Bojowald:2008ma} for a discussion of other corrections which may also need to be included in certain regimes).
The first 
originates from the spectra of quantum operators related to 
the inverse triad \cite{Thiemann:1996aw, inverse}, while the second  
arises from the use of holonomies as a
basic variable in the quantisation scheme \cite{rho2}.  
 One point to note here is that in the regime of very small scale factor,
the discrete structure of space is so strong that any inhomogeneous 
configuration would be far from being isotropic. Therefore, despite 
being able to write the effective perturbation equations for a purely
isotropic scenario in this regime, one should keep in mind that in 
reality care needs to be taken regarding the cosmological interpretation 
of perturbations and their evolution \cite{Bojowald:2008gw, Bojowald:2004ax}.

The two modifications 
have rather different origins, yet both can give 
rise to  
a period of super-inflation during which the Hubble rate rapidly increases, rather than
remaining nearly constant as is the case during standard slow-roll inflation. Since the
relative status of the two modifications discussed is at
present unclear, most works have taken the pragmatic 
approach of studying the dynamics when each of the
modifications is considered separately, but not including both sets of
modifications simultaneously. 

In earlier work \cite{Mulryne:2006cz, Copeland:2007qt} , we explored
super-inflationary scalar field models in the presence of each set of 
modifications in turn. We worked 
in an approximation which neglected back-reaction, and 
by studying scaling, power-law solutions we derived the form of the potential 
required in each case for perturbations to the scalar field, produced 
during the super-inflationary phase, 
to have a scale invariant spectrum.
Moreover, while the period of super-inflation in each case must 
be short lived, we showed that 
this is not a barrier for these scenarios since only a small number of 
e-folds of super-inflationary evolution is required to 
solve the horizon problem. In that work 
we did not, however, consider the spectrum of tensor 
perturbations produced by these scenarios, and this is the 
question to which we now turn. This question is particularly timely since 
the effect of the two kinds of modifications on the evolution of tensor perturbations in LQC 
has only recently been derived \cite{Bojowald:2007cd}.  The modifications have already been 
utilised to calculate  both corrections to the gravitational wave spectrum if standard 
inflation occurred in the presence of LQC corrections \cite{Barrau:2008hi} 
and the gravitational wave spectra from the `big bounce' \cite{Mielczarek:2008pf} which can occur in LQC.  

 Similar calculations have already been carried out and reported in the literature 
for different types of cosmologies. In particular, gravitational wave perturbations for 
ekpyrotic models of a collapsing universe \cite{Boyle:2003km},
and for phantom superinflation scenarios \cite{Piao:2006jz} have been computed. 
Despite being based on different physical concepts and behaviours, both
cosmologies predict a strongly blue tilted spectrum of tensor perturbations,
and since these have not yet been observed, it suggests that in both scenarios, the 
amplitude of these fluctuations are suppressed on large scales by many orders of magnitude
compared to those predicted by standard inflation. This is not unexpected, as it has been pointed out in Ref.~\cite{Lidsey:2004xd} that a duality exists  
between the ekpyrotic collapse and the dynamics of a universe sourced by a phantom field. Moreover, a scale factor duality maps the ekpyrotic collapse onto the superinflationary scaling solutions in LQC \cite{Lidsey:2004uz}.

Given the connections found between LQC and the ekpyrotic and phantom scenarios, one might expect that LQC also predicts a large blue tilted spectrum for the tensor perturbations. However, as the connections are at the level of the background equations and since LQC corrections also arise in the evolution equations of tensor perturbations themselves, this expectation may not be realised and a careful analysis is necessary to confirm it. We will see below that these corrections do not spoil this conjecture after all.

The structure of the paper is as follows. In Section \ref{InverseVolume}, we discuss the 
inverse triad modifications. First we review the background dynamics which give rise to a scale invariant 
spectrum of scalar field perturbations, and then we consider the evolution of tensor perturbations 
in this setting calculating their spectrum. We then repeat the exercise for holonomy corrections 
in section \ref{Holonomy}. Finally we conclude in section \ref{Discussion}.

\section{Tensor dynamics with inverse triad corrections}
\label{InverseVolume}

We first consider the cosmological equations which follow from including modifications 
associated with the inverse triad in LQC. This introduces two functions into the 
dynamics, $D_{j,l}(a)$ and $S_j(a)$. 
These functions arise because of the
presence of powers of the inverse scale factor in the
Hamiltonian constraint for an isotropic and
homogeneous universe.  A full discussion of the origin of these terms
can be found in Ref.~\cite{Bojowald:2002ny,Vandersloot:2005kh} (a summary can be found in appendix B of 
Ref.~\cite{Magueijo:2007wf}), but here we simply state their basic
properties. We note that we are implicitly considering either positively curved or topologically 
compact models. This ensures that the size of the fiducial cell does not 
enter in the equations of motion. $D$ and $S$ are both functions of 
the scale factor, and their
form changes depending on the values of two ambiguity parameters: $l$
which takes values in the range $0<l<1$, and $j$ which takes half
integer values. When the scale factor approaches zero, $D$ and $S$
also approach zero, whereas as $a$ increases above the critical value
$a_\star$, which depends on $j$, they both tend to unity. 
As noted in the introduction, for small values of the scale factor $a \ll a_\star$, the assumption of isotropy may be violated.
Whilst recognising the careful attention this regime deserves, and acknowledging more 
work is required to clarify the validity of the dynamical equations we use together
with the assumption of isotropy, we hope to pave the way
by demonstrating a method of calculation which can easily be employed once
new light is shed on currently uncertain sections of the theory.

The unperturbed isotropic equations of motion take the form of a modified Friedmann equation
\be
\label{FriedD}
\mathcal{H}^2 \equiv \left (\frac{a'}{a}\right)^2 = \frac{\kappa^2}{3} S \left( \frac{\phi'^2}{2D}+ a^2 V(\phi) \right)\,~,
\ee
where a dash represents differentiation with respect to conformal time and we have assumed that any curvature contribution has become subdominant, and the scalar field equation is
\be
\label{ScalarD}
\phi'' + 2 \mathcal{H} \left( 1- \frac{1}{2} \frac {d\ln D}{d \ln a} \right) {\phi'} + a^2 D V_{,\phi} =0\,.
\ee
For convenience, from here on we will work in units in which $\kappa^2=8 \pi G=1$.

In this study our interest is in the evolution of tensor perturbations about this isotropic background.  Tensor 
perturbations are defined as the transverse and trace free part of the perturbed spatial metric, and represent 
gravitational wave perturbations propagating on the unperturbed background spacetime. 
They can be further decomposed into two polarisation modes represented by $\times$ and $+$, and 
in LQC, with inverse triad modifications included, the equation of motion for these modes has recently been derived to be 
\cite{Bojowald:2007cd}
\be
\label{tensorS}
h''_{\times, +}  + 2 \mathcal{H} \left [1-\frac{1}{2}\frac{d \ln S}{d \ln a} \right ]  h'_{\times, +} - S^2 \nabla^2 h_{\times, +} = 0\,~.
\ee
$h_{\times, +}$ is a tensorial quantity, but from here on to avoid clutter 
we will drop the $\times$ and $+$ subscripts, and take $h$ to represent the 
magnitude of one of the polarisation modes, but always keeping in mind that both modes are present. 

\subsection{The background power law solution and scale invariant scalar field dynamics}

In Refs.~\cite{Lidsey:2004uz, Copeland:2007qt}, it was shown that in the regime $a \ll a_\star$ there exists a power law 
solution to the equations of motion (\ref{FriedD})--(\ref{ScalarD}) which is a stable attractor to 
the dynamics. In this regime $D(a) \approx D_{\star} a^n$, with $D_{\star} =
\left( 3/(3+2l) \right)^{3/2(1-l)} \, a_{\star} ^{-n}$ and $n = 3(3-l)/(1-l)$
takes values in the range $9<n<\infty$.
The function $S(a)$ may be similarly 
approximated by $S(a) \approx S_\star a^r$, where $S_\star = (3/2) a_\star^{-r}$ 
and $r = 3$, though we keep  $r$ arbitrary in our calculations for generality.
For $a \gg a_\star$ , $S_\star \approx D_\star \approx 1$
and $r = n = 0$.
The power law solution for $a \ll a_\star$ exists for negative 
power law potentials of the form $V=V_0 \, \phi^\beta $ and gives rise to the dynamics
\begin{eqnarray}
\label{atau}
a(\tau) &=& A(- \tau)^p \,,
\\
\label{phiprime}
\phi'(\tau) &=& -\sqrt{p \,\frac{2+(r+2)p}{\alpha}}\, \sqrt{\frac{D}{S}} \,\frac{1}{\tau} \,, 
\\
\label{Vtau}
V(\tau) &=& \left(3-\frac{2+(r+2)p}{2\alpha}\right) \,\frac{1}{ S (a\tau)^2}  \,,
\end{eqnarray}
where for an expanding universe $\tau$ is negative and increasing
towards zero. $A$ is an arbitrary normalisation 
constant, $\alpha=1-n/6$ with $\beta$ and $p$ being related through
\be
\beta = -\frac{2}{p} \, \frac{2+(r+2)p}{n-r} \,.
\ee
One can expand about this solution in terms of fast roll parameters, in order to generalise the potentials 
which can be considered \cite{Copeland:2007qt}, but here for simplicity we will consider only this exact solution.

For $-1<p < 0 $ the solution given above represents a universe undergoing 
super-inflationary evolution during which $H=\dot{a}/a$ (a dot denotes differentiation with respect to 
cosmic time), the Hubble rate, increases.  A particularly 
interesting case then occurs as $p$ tends to zero from below.  This represents a universe 
 in which the scale factor is almost constant, but $H$ increases rapidly. 
 Considering scalar field perturbations about the background field, it was shown in Refs. \cite{Mulryne:2006cz, Copeland:2007qt}, that the 
 spectrum of scalar field perturbations  attains scale invariance in this limit. Moreover, because $H$ 
 increases so rapidly,
  the horizon problem is solved during this phase with only a small number of $e$-folds required.  
This raises the intriguing possibility that these perturbations could be responsible for the observed 
CMB anisotropies and hence for structure in the universe.  If this were the case,  
no period of standard inflation in which $H$ remains nearly constant for roughly $60$ 
$e$-folds of expansion would be required (where number of $e$-folds is defined as $N=\ln(a/a_{\rm i})$). A natural, and indeed important question is: `what is the spectrum of primordial tensor 
perturbations which accompanies 
this scale invariant spectrum of scalar field perturbations?'. 
This is the question to which we now turn.

\subsection{The primordial spectrum of tensor perturbations}

To calculate the spectrum of gravitational waves produced during super-inflation we 
follow the standard procedure. Noting that $2 a^2 h$ is canonically 
conjugate to $h'/S $ such that \cite{Bojowald:2007cd}
\be
\left \{ \frac{1}{2 S} h'(t,x), -\frac{a^2}{2} \, h(t,y) \right \} =  \delta^3(x,y)
\ee 
the system is quantised by promoting 
$h$ and $h'$ to operators and the Poisson brackets to commutators. We have 
\be
\label{com}
\left [\hat{h}',\hat{h} \right] = -4 i \frac{S}{a^2} \delta^3(x,y)\, .
\ee
$h$ is then decomposed into Fourier modes
\be
\label{hoperator}
\hat h= \int \frac{d^3k}{(2 \pi )^{3/2}} \left [ h_k(\tau ) \hat
a_{\rm \bf k} e^{i{\rm \bf k \cdot x}} + h_k^*(\tau ) \hat a_{\rm \bf k}^{\dagger} e^{-i{\rm \bf k \cdot x}} \right ]\, ~,
\ee
where, considering Eq.~(\ref{tensorS}), we see that 
each mode $h_k$ obeys the evolution equation
\be
\label{tensorSk}
h''_k  + 2 \mathcal{H} \left (1-\frac{1}{2}\frac{d \ln S}{d \ln a} \right )  h'_k +S^2 k^2 h_k = 0 \,~.
\ee

The power spectrum for one polarisation  state of 
tensorial fluctuations is given by the standard expression
\be
\label{power}
{\cal P}_h=\frac{k^3}{2 \pi^2} \, \left | h_k \right |^2 \,~.
\ee

In order to evaluate Eq.~(\ref{power}) 
we must solve Eq. (\ref{tensorSk}). 
Considering the power law solution for the 
regime $a \ll a_\star$ (Eq.~(\ref{atau})), and using the form of $S$ in this regime, we find that  
Eq.~(\ref{tensorSk}) becomes
\be
\label{tensorSkt}
h''_k  + 2 \,\frac{p}{\tau}  \left (1-\frac{r}{2} \right )  h'_k + k^2 S_{\star}^2 A^{2r} (-\tau)^{2 r p} h_k = 0 \,~.
\ee
which admits the exact solution
\be
\label{hSol1}
h_k(\tau) = \frac{S^{1/2}}{ \mathcal{H}^{1/2} a} \sqrt{\frac{-p \,\pi}{1+rp}} \, H_\nu^{(1)}(x) \,,
\ee
where 
\be
\nu = \frac{1+p(r-2)}{2(1+pr)} \,, \hspace{1cm}
x = \frac{-p S k}{(1+pr) {\cal H}} \,,
\ee
and we have normalised the solution such that $a_{\rm \bf k}$ and 
$a^{\dagger}_{\rm \bf k}$ satisfy the usual raising and lowering operator algebra while 
$\hat{h}$ and $\hat{h}'$ satisfy 
the commutation algebra (\ref{com}), such that only the forward moving 
solution is selected in the asymptotic past (the adiabatic vacuum).
The solution 
(\ref{hSol1}) has the expected behaviour that each $k$ mode begins in an oscillatory 
state where normalisation occurs, and evolves into a non-oscillatory  state once each 
given $k$ mode crosses a suitably defined horizon.  From the form of (\ref{hSol1}), it is clear that 
for tensor modes horizon crossing occurs when $k \approx -a H /pS$.  
A given mode can only be considered 
to become a classical perturbation once it crosses this horizon.
 
Taking the solution (\ref{hSol1}) and employing (\ref{power}) we find that 
\be
\label{spectrum1}
{\cal P}_{h} = \frac{2^{2(\nu-1)}}{\pi^2} \frac{\Gamma(\nu)^2}{\Gamma(3/2)^2} \left(\frac{k_0}{k_{\rm e}}\right)^{3-2\nu} \left(\frac{k}{k_0}\right)^{3-2\nu} H_{\rm e}^2 \,,
\ee
We have evaluated (\ref{hSol1}) in the limit where the modes 
are outside the defined horizon (i.e. $x \to 0$), and have accounted for 
both polarisation states. The mode $k_0$ that corresponds to the largest scales on the 
CMB and $k_{\rm e}$ is defined as the last mode to cross the horizon 
at the end of super-inflation ($k_{\rm e} = \mathcal{H}_{\rm e}/S_{\rm e} \approx \mathcal{H}_{\rm e}$). 

A couple of observations are in order. The 
first is that in the limit of interest, $p\to 0$, the spectrum is blue tilted with a tensor 
spectral index given by $n_t =2$, where
$n_t$ 
is defined, as usual,  by $P_h \propto k^{n_t}$. This implies that on large scales the spectrum is 
hugely suppressed. The 
second important point is that the magnitude of the spectrum is fixed by the Hubble rate at the end of inflation. 
This also fixes the magnitude of scalar perturbations and ultimately has to be normalised 
such that the scalar perturbations have the correct magnitude to account for the CMB anisotropies.  In the 
scenario at hand such a normalisation is difficult 
to determine since the scalar field perturbations (which are close to scale 
invariant) must be related to curvature perturbations (as discussed at length in Refs. \cite{Copeland:2007qt}), and this step 
cannot be performed at present. Nevertheless in the following section we will make the reasonable assumption 
that $H_{\rm e} \approx 10^{-6}$ corresponding to the GUT scale
in order to calculate the present-day spectrum of gravitational waves produced by this super-inflationary scenario.  Furthermore, we will see that the conclusion - that the spectrum is unobservably small - is insensitive to the 
choice of normalisation within reasonable bounds.

\subsection{The present-day spectrum}

\begin{figure}[]
\includegraphics[width = 8.5cm]{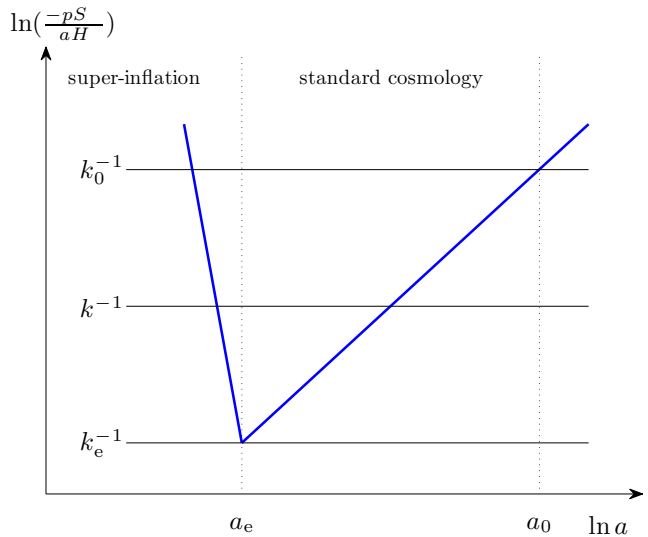}
\caption{\label{fig5} Schematic illustration of the evolution of modes $k$ which exit the horizon 
(thick solid line) during super-inflation and re-enter during the standard radiation or matter era. 
$k_0^{-1}$ is the scale corresponding to the size of the observable universe.}
\end{figure}

At the end of super-inflation all classical perturbations modes are outside the horizon. We will assume 
that reheating occurs instantaneously at the energy scale $H_{\rm e}$, and hence that 
the universe becomes radiation dominated at this point.  
Moreover we will assume the universe is classical after reheating, and any quantum 
corrections (similar to $D$ or $S$) are absent from the dynamics. From this point onwards 
modes will begin to re-enter the cosmological horizon. We schematically illustrate this dynamics in Fig.~\ref{fig5}.

To convert the primordial spectrum, Eq.~(\ref{spectrum1}), to the spectrum which would be observed today we employ the 
numerically obtained transfer function \cite{Turner:1996ck, Turner:1993vb}
\be
\label{transfer}
T(k) = \left ( \frac{k_0}{k} \right )^2 \left [ 1+ \frac{4}{3} \frac{k}{k_{\rm eq}} + \frac{5}{2} \left(\frac{k}{k_{\rm eq}}\right)^2 \right]^{1/2}\,,
\ee
where $k_0=H_0 a_0$, is again the $k$ mode corresponding to the largest scales today, and $k_{\rm eq}=a_{\rm eq} H_{\rm eq}$, where 
eq stands for quantities at radiation matter equality. 
This form arises from the observation that tensor modes outside the horizon are roughly time independent while 
modes inside decay as $a^{-1}$, together with the evolution of $H \propto a^{-2}$ during 
radiation domination and $H \propto a^{-3/2}$ during matter domination.  

Using this transfer function we can calculate the present day spectrum of tensor perturbations once 
we fix $H_{\rm e}$ and $k_{\rm e}/k_0$.  A sensible estimate for $H_{\rm e}$ is the GUT scale, while an absolute 
upper limit is given by the Planck scale. If we also make the assumption that as many modes exit the tensor 
horizon as scalar modes exit the scalar horizon, 
$k_0/k_{\rm e}$ can be 
fixed by the requirement that the horizon problem be solved.  
In Ref. \cite{Copeland:2007qt} we found that this required 
$k_0/k_{\rm e}\approx e^{-60}$.  

A useful physical quantity that can be used to express the present day 
spectrum of gravitational waves is $\Omega_{\rm gw}$, the gravitational 
wave energy per unit logarithmic wave number in units of the critical 
density, $\rho_{\rm crit}=3 H_{0}^2$, \cite{Boyle:2003km, Turner:1996ck}:

\be
\label{gw}
\Omega_{\rm gw}(k) \equiv \frac{k}{\rho_{\rm crit}} \, \frac{d\rho_{\rm gw}}{dk} = \frac{1}{6}\, \left(\frac{k}{k_0}\right)^2 \, T^2(k) \, {\cal P}_{h}\,.
\ee

\begin{figure}[]
\includegraphics[width = 8.5cm]{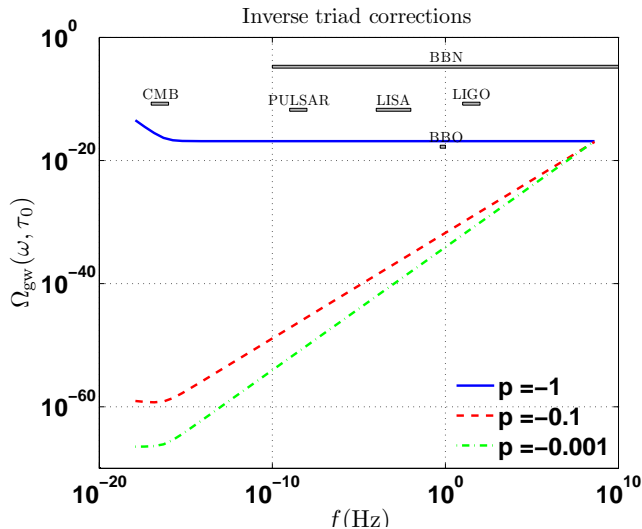}
\caption{\label{fig1} Predicted present abundance of gravitational waves produced 
during the super-inflationary phase of LQC with inverse volume corrections assuming 
that $H_{\rm e} = 10^{-6}$ in Planck units. The solid line corresponds to the standard inflation 
abundance ($n = r = 0$). Also indicated are the present bounds and future sensitivities.}
\end{figure}

Figure \ref{fig1} shows a plot of the present tensor abundance for a number of choices of the parameter $p$, 
together with the strongest observational constraints from current and future experiments searching for gravitational waves \cite{Observations}.

It can be verified that even the limiting value of $H_{\rm e} \approx 1$ (giving the largest spectrum 
possible), leads to an unobservably small spectrum of tensor perturbations. 

\section{Tensor dynamics with holonomy corrections}
\label{Holonomy}

We now turn our attention to the second set of effective equations, those which arise 
from considering that holonomies are the basic operators to be quantised in LQC.

The isotropic unperturbed dynamics is described by the Friedmann equation \cite{rho2}:
\be
\label{H^2_nonmin}
\mathcal{H}^2 = \frac{1}{3} \, a^2 \rho \, \left(1- \frac{\rho}{2\sigma}\right) \,,
\ee
which is modified from the classical equation by the inclusion of a $-\rho^2$ term,  where 
$\sigma = 3/ (2 \gamma^2)$ is a constant in an exactly isotropic model, and where 
$\gamma$ is the Barbero-Immirzi parameter \cite{Barbero:1994ap, Immirzi:1996di}. 
Once again we are assuming either a flat universe or that the curvature 
contribution can be safely neglected. Our interest here is 
in a scalar field dominated universe, and hence $\rho = \phi'^2/2 a^2 + V(\phi)$.
The scalar field equation of motion remains unchanged from its classical form
\be
\label{e.o.m_nonmin}
\phi''+2 \mathcal{H} \phi' + a^2 V_{,\phi} = 0 \,.
\ee

We stress that as we are studying inverse volume and quadratic 
corrections separately,  we do not include 
the $D$ and $S$ functions in the equations of motion.

The form of the evolution equation for tensor perturbations when holonomy corrections are included has recently 
been derived to be \cite{Bojowald:2007cd}
\be
\label{tensorRho}
h''_{\times, +}  + 2 \mathcal{H} h'_{\times, +} - \nabla^2 h_{\times, +} + T_Q h_{\times, +}= 2 \Pi_Q \,~,
\ee
where 
\begin{eqnarray}
\label{TPi}
T_Q &=& \frac{a^2}{3}  \frac{\rho^2}{2\sigma}\,,\\  
\Pi_Q &=& \frac{1}{2} \, \frac{\rho}{2\sigma} \left(\frac{a^2}{3}\, \rho - \phi'^2 \right) \, h_{\times,+} \,,
\end{eqnarray}
are quantum corrections to the classical dynamics that become unimportant when $\rho \ll 2 \sigma$.
From here on we again drop the $\times$ and $+$ subscripts as we did in the inverse triad case.

\subsection{Power-law solution and scale invariant scalar field perturbations}

We are interested in high density regimes where $\rho$ approaches the bounding  value of $2 \sigma$. In this case, the term within brackets of Eq.~(\ref{H^2_nonmin}) 
tends to zero and the behaviour of the 
equations alters significantly compared with the classical
behaviour. Indeed in this regime we have $\dot{H}>0$ and
for an expanding universe super-inflation takes place.  In our previous work \cite{Copeland:2007qt}, 
we showed that in this regime there exists an approximate power-law solution for a 
potential of the form $V = 2\sigma - U_0 e^{-\lambda\phi} $. 
Moreover, this solution is 
an attractor as we demonstrated both analytically and 
numerically \cite{Copeland:2007qt}. The full solution is given by
\begin{eqnarray}
\label{atau2}
a(\tau) &=& A(-\tau)^p \,,
\\
\label{phiprime2}
\phi'(\tau) &=&  \sqrt{-2p(p+1)} \, \frac{1}{\tau} \,,
\\
\label{Vtau2}
V(\tau) &=& 2\sigma - \frac{p(2p-1)}{(a\tau)^2}
\end{eqnarray}
where $A$ is an arbitrary normalisation constant. $\lambda$ and $p$ are related through
\be
\lambda^2 = - 2 \, \frac{p+1}{p} \,.
\ee
Once again $ -1< p < 0$ corresponds to a universe undergoing super-inflation, 
and moreover the limit $p \to 0$ from below leads to a scale invariant spectrum of scalar field perturbations. Furthermore, 
as was the case for the super-inflationary solution we studied in the presence of inverse volume corrections, 
only a small number of $e$-folds are required to solve the horizon 
problem.  In our previous work we generalised the form of the potential that could be considered 
by expanding about this solution, 
but for simplicity in this work we will consider only this exact form.

We now turn our attention to the question of what spectrum of tensor perturbations accompanies
the scale invariant spectrum of scalar field perturbation in this version of super-inflation.

\subsection{The primordial spectrum of tensor perturbations}

We again follow the standard procedure for calculating the 
spectrum of tensor perturbations.  In this case $a^2 h$ is canonically conjugate to $h'$. 
To quantise the system $h$ and $h'$ are promoted to operators and Fourier decomposed according to 
Eq.~(\ref{hoperator}). Each $h_k$ mode 
satisfies the evolution equation
\be
\label{tensorSk2}
h_k'' + 2 \frac{a'}{a} \, h_k' + \left[ k^2 - \frac{2p(p+1)}{\tau^2} \right] h_k = 0 \,,
\ee
during the scaling solution, where we have used $\rho \approx 2\sigma$ and Eq.~(\ref{phiprime2}). 

\begin{figure}[]
\includegraphics[width = 8.5cm]{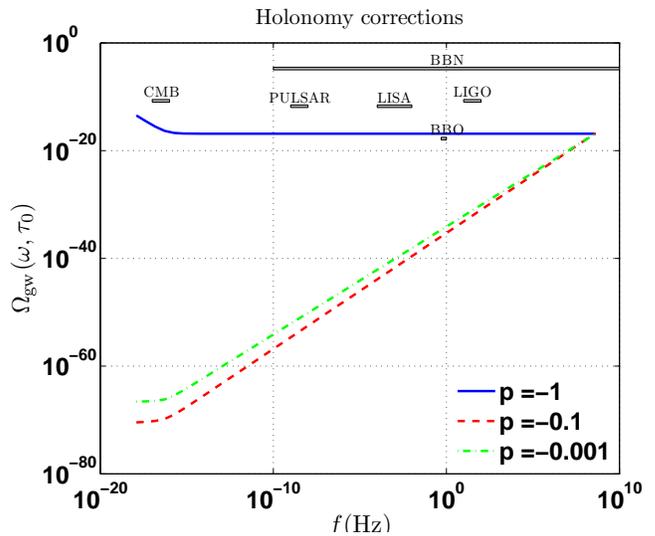}
\caption{\label{fig2} Predicted present abundance of gravitational waves 
produced during the super-inflationary phase of LQC with holonomy corrections.
The solid line corresponds to the standard inflation abundance. 
Also indicated are the present bounds and future sensitivities.}
\end{figure}

The solution to Eq.~(\ref{tensorSk2}),  is given by
\be
\label{hSol2}
h_k(\tau) = \frac{1}{ \mathcal{H}^{1/2} a} \sqrt{-p \,\pi} \, H_\nu^{(1)}(-k\tau) \,,
\ee
where
\be
\nu = \frac{1}{2} \sqrt{1+4p+12p^2} \,,
\ee
where we have normalised the solution by the requirement that $a_{\rm \bf k}$ and $a^{\dagger}_{\rm \bf k}$ 
satisfy the usual raising and lowering operator algebra while $a^2 \hat{h}$ and $\hat{h}'$ satisfy 
their commutation algebra, with only the forward moving 
solution being selected in the asymptotic past (selecting the adiabatic vacuum). 
Finally, utilising Eq.~(\ref{power}) and evaluating this expression in the limit that the modes 
are outside the horizon ($k < -\mathcal{H}/p$ in this case), leads us to the same expression 
for the primordial power spectrum (\ref{spectrum1}). 
The spectral index can clearly be seen to be given by $n_t=2$ in the limit $p\to 0$, 
and the amplitude fixed by $H_{\rm e}$.  We note here that in cases where
the dynamics of tensor perturbations could be represented by the standard equation
of motion and the evolution of the scale factor is given by Eq. (\ref{atau2}), having 
$p \to 0$ will inevitably result in $n_t =2$. This has indeed been shown to be true 
explicitly for the collapsing ekpyrotic scenario \cite{Boyle:2003km} and for the evolution of a
universe sourced by a phantom field \cite{Piao:2006jz}.
At this point, making some reasonable assumptions, we 
can proceed, as we did for the solution in the inverse triad case, to calculate the present day spectrum 
of gravitational wave perturbations using the transfer function Eq.~(\ref{transfer}). 
We show the abundance of tensors in Fig.~\ref{fig2} and note that the result is very similar to the inverse volume case.

\section{Discussion}
\label{Discussion}

We conclude that super-inflation in both versions of the corrections in LQC predicts a
strong blue spectrum of gravitational waves, hence, their abundance is strongly suppressed on the large scales and it is 
many orders of magnitude smaller than the value  predicted in the standard inflationary scenario.

It is now important to discuss our results in the light of other investigations in the literature. 
In particular, it is claimed in Ref.~\cite{Mielczarek:2007wc} that the abundance of gravitational 
waves generated during super-inflation under inverse volume corrections is above the current bounds. 
We note however that the authors did not consider the evolution of the background in a scaling 
solution and the $S(a)$ correction was not used. Moreover, the full expression for the tensor 
perturbations was obtained more recently \cite{Bojowald:2007cd} and therefore, it was not used in that work.
In a more recent analysis \cite{Barrau:2008hi} focusing on the holonomy corrections, 
it is also found, like in our work, that the spectrum of gravitational waves must be blue. 
However, a scaling solution was not used and the expansion of the universe was assumed 
to be close to de Sitter, hence, a direct comparison with our work is in fact not possible.

 One needs to be cautious about making general conclusions in the context of LQC, 
as the theory is currently far from complete. As mentioned previously, the introduction 
of inhomogeneities in the small $a$ regime is likely to break the assumption of an 
isotropic universe. Attempts are being made to gain better understanding of this region 
\cite{Bojowald:2008gw}.  
Moreover, there is the possibility that higher order perturbative correction terms could play a role, or that quantum backreaction might significantly modify the background dynamics \cite{Bojowald:2008ma}.

Our work on super-inflationary scenarios in LQC also has a number of other possible drawbacks. We have treated the inverse triad and holonomy corrections separately while they should be dealt with together in a realistic set up. Though the fact that 
both sets of corrections lead to such similar phenomenology gives us some reassurance that combining them could lead to qualitatively similar results. 
Furthermore, so far we have not investigated the evolution of scalar metric perturbations. The derivation of the full equations for these is still in progress \cite{metricPerts} and although these equations are not required for the calculation of tensor perturbations presented here, they  are required to understand how the scale-invariant scalar field perturbation is related to the observed curvature perturbation. Finally, we should mention recent work where it has been shown that the behaviour of the LQC equations with holonomy corrections in the presence of negative 
exponential scalar field potentials leads to sudden singularities where the Hubble rate is bounded, 
but the Ricci curvature scalar diverges \cite{Cailleteau:2008wu}. Given that for the case 
of holonomy corrections our superinflation scenario 
requires a scalar field potential with a negative exponential part, this serious problem needs 
to be avoided.  In the scenarios we consider, however, the potentials only need to be of the form 
which gives rise to the power-law behaviour while superinflation is taking place. After this phase of evolution the potential can change in form.  
For example, any potential which tended to zero after the field evolved through the region which 
gives rise to the 
super-inflation phase would avoid the sudden singularity problem.

Despite the ambiguities and uncertainties in the theory, if we were to make observational claims for the scaling solutions we have considered in the current state of LQC, it would be that if gravitational waves are observed, they would rule out this scenario of superinflation in LQC sas it stands.

\begin{acknowledgments}
DJM is supported by the Centre for Theoretical Cosmology, Cambridge, NJN by STFC and MS by a University of Nottingham 
bursary.  We would like to thank Martin Bojowald and Parampreet Singh for helpful discussions.
\end{acknowledgments}

\end{document}